\documentclass[aps,prb,reprint,twocolumn,superscriptaddress,showpacs,amssymb]{revtex4-2}
\usepackage{amsmath, nccmath}
\usepackage{amssymb}
\usepackage{bm}
\usepackage{braket}
\usepackage{natbib}
\usepackage{leftidx}
\usepackage{graphicx}    
\usepackage{verbatim}   
\usepackage{color}      
\usepackage{subfigure}  
\usepackage{hyperref}   
\usepackage{dcolumn}    
\usepackage{textcomp}
\usepackage{float}
\usepackage{threeparttable}
\usepackage{titlecaps}
\usepackage{multirow}
\usepackage{lipsum}
\hypersetup{
	colorlinks=true,
	citecolor=blue,
	filecolor=black,
	linkcolor=blue,
	urlcolor=cyan
}
\usepackage{mathastext}
\usepackage{mathptmx}
\usepackage{enumitem}
\DeclareGraphicsExtensions{.png,.pdf,.tif}
\usepackage{pict2e}
\usepackage[dvipsnames]{xcolor}
\usepackage[normalem]{ulem}

\begin{document}

\title{Investigation of the Anomalous and Topological Hall Effects in Layered Monoclinic Ferromagnet Cr$_{2.76}$Te$_4$}
\author{Shubham Purwar}
\affiliation{Department of Condensed Matter and Materials Physics, S. N. Bose National Centre for Basic Sciences, Kolkata, West Bengal, India, 700106.}
\author{Achintya Low}
\affiliation{Department of Condensed Matter and Materials Physics, S. N. Bose National Centre for Basic Sciences, Kolkata, West Bengal, India, 700106.}
\author{Anumita Bose}
\affiliation{Solid State and Structural Chemistry Unit, Indian Institute of Science, Bengaluru 560012, India.}
\author{Awadhesh Narayan}
\affiliation{Solid State and Structural Chemistry Unit, Indian Institute of Science, Bengaluru 560012, India.}
\author{S.\ Thirupathaiah}
\email{setti@bose.res.in}
\affiliation{Department of Condensed Matter and Materials Physics, S. N. Bose National Centre for Basic Sciences, Kolkata, West Bengal, India, 700106.}

%

\date{\today}

\begin{abstract}
We studied the electrical transport, Hall effect, and magnetic properties of monoclinic layered ferromagnet Cr$_{2.76}$Te$_4$. Our studies demonstrate Cr$_{2.76}$Te$_4$ to be a soft ferromagnet with strong magnetocrystalline anisotropy. Below 50 K, the system shows an antiferromagnetic-like transition. Interestingly, between 50 and 150 K, we observe fluctuating magnetic moments between in-plane and out-of-plane orientations,  leading to non-coplanar spin structure. On the other hand, the electrical resistivity data suggest it to be metallic throughout the measured temperature range, except a $kink$ at around 50 K due to AFM ordering. The Rhodes-Wohlfarth ratio  $\frac{\mu_{eff}}{\mu_{s}}=1.89 (>1)$ calculated from our magnetic studies confirms that Cr$_{2.76}$Te$_4$ is an itinerant ferromagnet. Large anomalous Hall effect has been observed due to the skew-scattering of impurities and the topological Hall effect has been observed due to non-coplanar spin-structure in the presence of strong magnetocrystalline anisotropy. We examined the mechanism of anomalous Hall effect by employing the first principles calculations.
\end{abstract}

\maketitle

\section{Introduction}
Two-dimensional (2D) magnetic materials with topological  properties~\cite{augustin2021properties,PhysRevX.11.031047,choudhary2020computational} have sparked significant research attention recently due to their potential applications in spintronics and magnetic storage devices~\cite{nagaosa2013topological,schaibley2016valleytronics,fert2017magnetic}.
Importantly, these are the van der Waals (vdW) magnets possessing peculiar magnetic properties with strong magnetocrystalline anisotropy (MCA)~\cite{ gong2017discovery,seyler2018ligand,fei2018two}.  In general, the Heisenberg-type ferromagnet does not exist with long-range magnetic ordering at a finite temperature in the 2D limit due to dominant thermal fluctuations~\cite{PhysRevLett.17.1133}. However, the strong magnetic anisotropy that usually present in the low-dimensional materials  can stabilize the long-range magnetic ordering to become a 2D Ising-type ferromagnet~\cite{Huang2017}. Till date many 2D ferromagnets have been discovered experimentally~\cite{Burch2018, Gong2019}, but only a few of them can show the topological signatures such as the topological Hall effect (THE) or skyrmion lattice. For instance, the recent microscopic studies on Cr$_2$Ge$_2$Te$_6$~\cite{han2019topological}, Fe$_3$GeTe$_2$~\cite{PhysRevB.100.134441}, and Fe$_5$GeTe$_2$~\cite{PhysRevB.105.014426,schmitt2022skyrmionic} demonstrated topological magnetic structure in the form of skymion bubbles in their low-dimensional form.

On the other hand, soon after predicting the layered Cr$_x$Te$_y$-type systems as potential candidates to realize the 2D ferromagnetism in their bulk form~\cite{PhysRevMaterials.2.081001,PhysRevB.96.134410}, a variety of Cr$_x$Te$_y$ compounds were grown including CrTe~\cite{ETO200116}, Cr$_2$Te$_3$~\cite{wang2018ferromagnetic}, Cr$_3$Te$_4$~\cite{hessen1993hexakis}, and Cr$_5$Te$_8$~\cite{PhysRevB.100.024434}. Interestingly, all these systems are formed by the alternative stacking of Cr-full (CrTe$_2$-layer) and Cr-vacant (intercalated Cr-layer) layers along either $\it{a}$-axis or $\it{c}$-axis~\cite{Dijkstra_1989}. Thus, the Cr concentration plays a critical role in forming the crystal structure, magnetic, and transport properties. The compounds like Cr$_5$Te$_8$, Cr$_2$Te$_3$, and Cr$_3$Te$_4$ are reported to crystallize in monoclinic or trigonal structures, whereas Cr$_{1-x}$Te $(x<0.1)$ crystallizes in the hexagonal NiAs-type structures~\cite{IPSER1983265}. The electronic band structure calculations performed on CrTe, Cr$_2$Te$_3$, and Cr$_3$Te$_4$ suggest a strong out-of-plane Cr $3d$ $e_g$ orbital,  $d_{z^2}-d_{z^2}$,  overlapping along the $\it{c}$-axis to have relatively smaller nearest neighbor $Cr-Cr$ distance~\cite{Dijkstra_1989}. In addition, Cr$_5$Te$_8$~\cite{PhysRevB.100.024434}, Cr$_{1.2}$Te$_2$~\cite{huang2021possible}, Cr$_{0.87}$Te~\cite{liu2022magnetic} are known to show topological properties in the hexagonal phase.

In this work, we systematically investigate the electrical transport, Hall effect, and magnetic properties of monoclinic Cr$_{2.76}$Te$_4$ which is very close to the stoichiometric composition of Cr$_3$Te$_4$. We observe that the easy-axis of magnetization is parallel to the $\it{bc}$-plane, leading to strong magnetocrystalline anisotropy. Below 50 K, the system shows antiferromagnetic-like transition. In addition, we find fluctuating Cr magnetic moments between in-plane and out-of-plane directions within the temperature range of 50 and 150 K. Electrical resistivity data suggest Cr$_{2.76}$Te$_4$ to  metallic throughout the measured temperature range with a $kink$ at around 50 K due to AFM ordering. Our studies clearly point Cr$_{2.76}$Te$_4$ to an itinerant ferromagnet. Magnetotransport  measurements demonstrate large anomalous Hall effect (AHE) and topological Hall effect (THE) in this system. First-principles calculations point to an intrinsic AHE due to non-zero Berry curvature near the Fermi level, while experimentally it is found to be an extrinsic AHE due to the skew-scattering~\cite{Smit1958}.

 \begin{figure}[t]
    \centering
    \includegraphics[width=\linewidth]{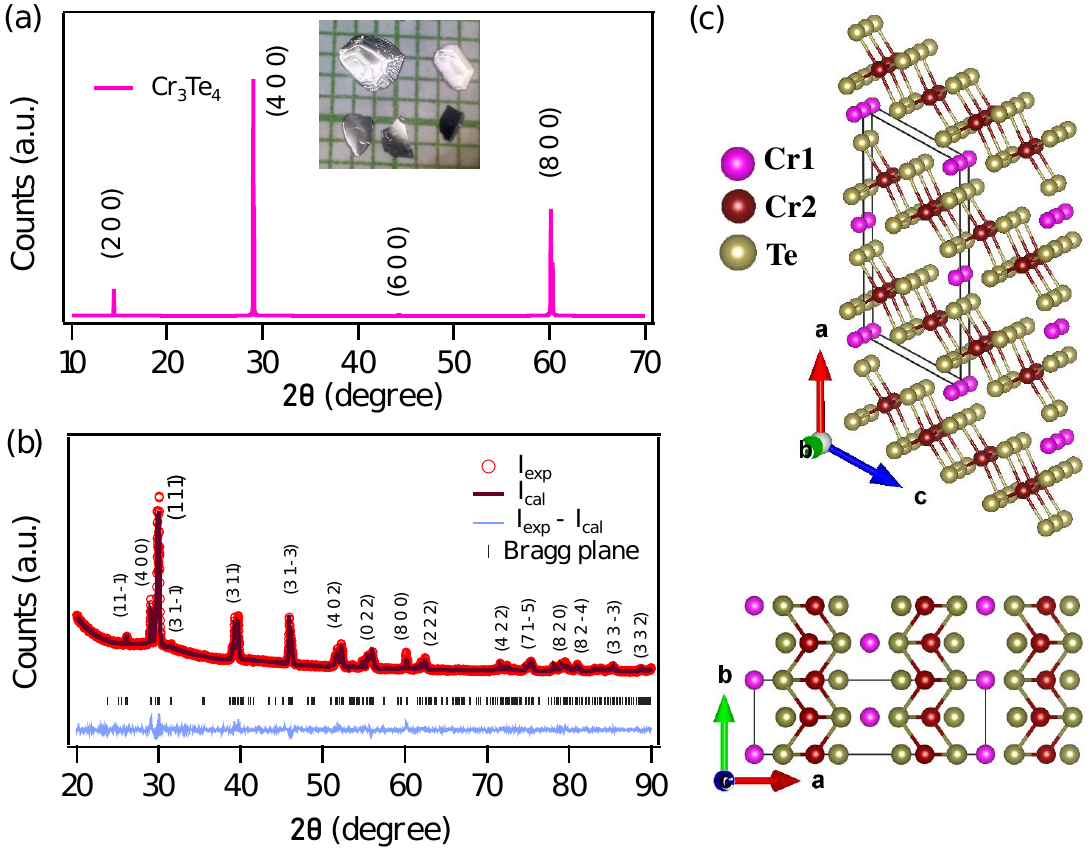}
    \caption{(a) XRD pattern from Cr$_{2.76}$Te$_4$ single crystals. Inset in (a) shows the photographic image of the single crystals. (b) X-ray diffraction pattern from the crushed Cr$_{2.76}$Te$_4$ single crystals, overlapped with Rietveld refinement. (c) Schematic crystal structure of Cr$_{2.76}$Te$_4$ obtained from the Rietveld refinement.}
    \label{1}
\end{figure}

\section{Experimental details}
High quality single crystals of Cr$_{2.76}$Te$_4$ were grown by the chemical vapor transport (CVT) technique with iodine as a transport agent as per the procedure described earlier~\cite{hashimoto1971magnetic}. Excess Iodine present on the crystals was removed by washing with ethanol several times and dried under vacuum. The as-grown single crystals were large in size ($3\times2$ mm$^2$), were looking shiny, and easily cleavable in the $\it{bc}$-plane. Photographic image of typical single crystals is shown in the inset of Fig.~\ref{1}(a). Crystal structural and phase purity of the single crystals were identified by the X-ray diffraction (XRD) technique using Rigaku X-ray diffractometer (SmartLab, 9kW) with Cu K$_\alpha$ radiation of wavelength 1.54059 \AA. Compositional analysis of the single crystals was done using the energy dispersive X-ray spectroscopy (EDS of EDAX). Magnetic and transport studies were carried out on the physical property measurement system (9 Tesla-PPMS, DynaCool, Quantum Design). See Supplemental Material for more discussion on the chemical composition of the studied system~\cite{suppl}.  Electrical resistivity and Hall measurements were performed in the standard four-probe method. To eliminate the longitudinal magnetoresistance contribution due to voltage probe misalignment, the Hall resistance was calculated as $\rho_{\it{yz}}$(H)=$[\rho_{\it{yz}}(+H)-\rho_{\it{yz}}(-H)]/2$.

\begin{figure}[htbp]
    \centering
    \includegraphics[width=\linewidth]{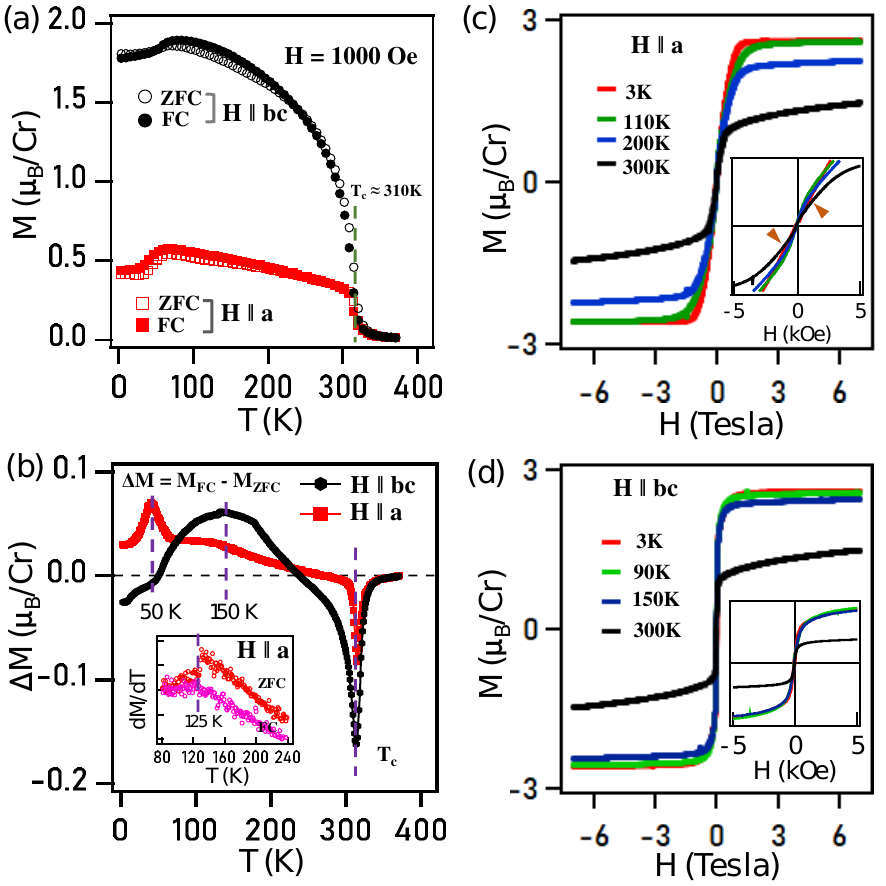}
    \caption{(a) Temperature dependent magnetization $M(T)$ measured under ZFC and FC modes with a magnetic field $H$=1000 Oe for $H\parallel \it{a}$ and  $H\parallel \it{bc}$. (b) Variation of magnetization $\Delta$M=(M$_{FC}$-M$_{FC}$) plotted as a function of temperature. Inset in (b) shows first derivative of magnetization with respect to the temperature (dM/dT) of the data shown in (a) for $H\parallel \it{a}$.  (c) and (d) Field dependent magnetization $M(H)$ measured at different temperatures for $H\parallel \it{a}$ and  $H\parallel \it{bc}$, respectively.}
    \label{2}
\end{figure}

\section{Density Functional Theory Calculations}
We have performed density functional theory (DFT) calculations using the Quantum Espresso package~\cite{Giannozzi2009, Giannozzi2017}. We used fully relativistic pseudopotentials in order to include the spin-orbit interaction. Generalized gradient approximation was
considered based on Perdew-Burke-Ernzerhof implementation~\cite{Perdew1996} within the projector augmented wave (PAW) basis~\cite{Bloechl1994}. For wave function and charge density expansions, cutoff values of 50 Ry and 300 Ry were chosen, respectively. For the self-consistent calculation, a 7$\times$7$\times$7 Monkhorst-Pack grid was used~\cite{Monkhorst1976}. In order to consider the van der Waals forces, Semi-empirical Grimme’s DFT-D2 correction~\cite{Grimme2006} was included. We further constructed tight-binding model based on the maximally localized Wannier functions using the wannier90 code~\cite{Mostofi2008}, with Cr $3d$, Cr $3s$, Te $5p$, and Te $5s$ orbitals as the basis. Then utilizing the as obtained tight-binding model, we calculated Berry curvature along the high symmetry directions using Kubo formula~\cite{Thouless1982} encoded in wannier90 code~\cite{Mostofi2008}. We have calculated intrinsic anomalous
Hall conductivity (AHC) by integrating the $\it{x}$-component of Berry curvature over the entire BZ using WannierTools code~\cite{Wu2018}.

\begin{figure*}[t]
    \centering
    \includegraphics[width=0.85\linewidth]{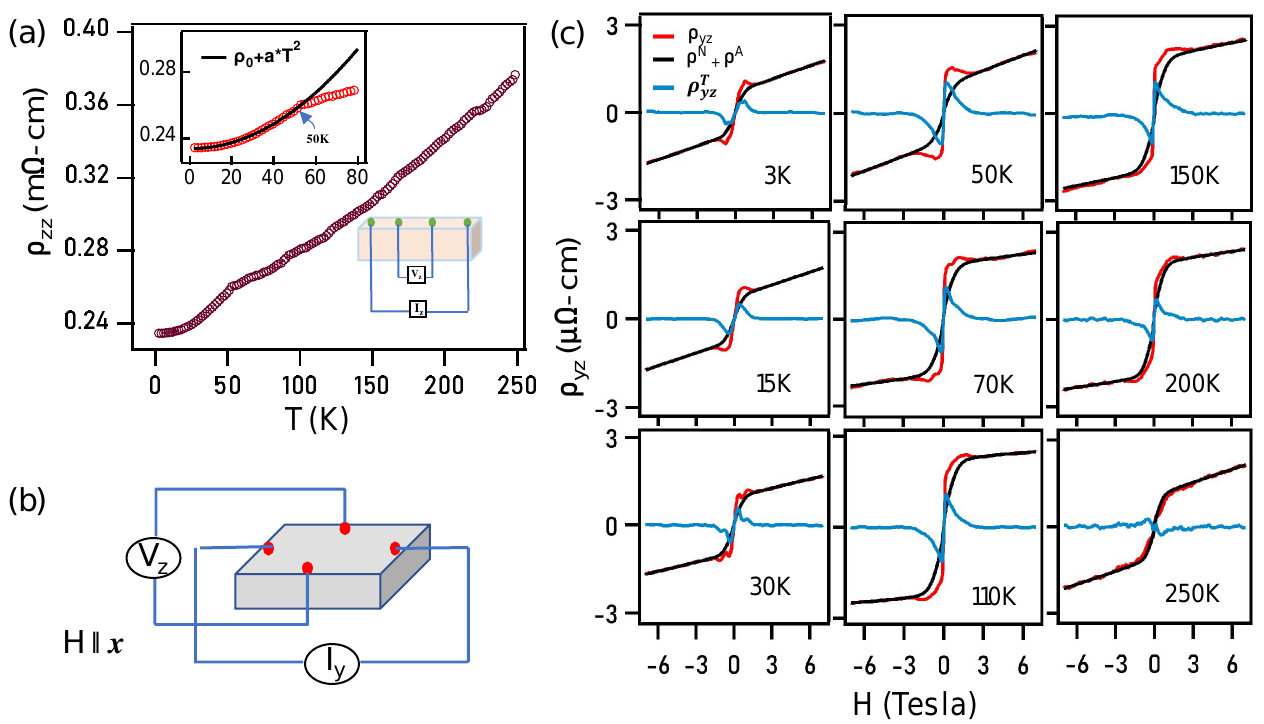}
    \caption{(a) Temperature dependent longitudinal resistivity $\rho_{\it{zz}}$(T). Top-inset in (a) shows low temperature resistivity fitted by $\rho$(T) = $\rho_0$ + bT$^2$ and bottom-inset in (a) shows schematic diagram of linear-four-probe contacts. (b) Hall measuring geometry is shown schematically. (c) Hall resistivity ($\rho_{\it{yz}}$) plotted as a function of magnetic field measured at various temperatures.  In (c), Red curves represent the experimental data of total Hall resistivity ($\rho_{\it{yz}}$), black curves represent the contributions from the normal and anomalous Hall resistivities ($\rho^N+\rho^A$), and Blue curves represent the topological Hall resistivity ($\rho^{T}_{\it{yz}}$). See the text for more details.}
    \label{3}
\end{figure*}

 \begin{figure}[b]
    \centering
    \includegraphics[width=0.95\linewidth]{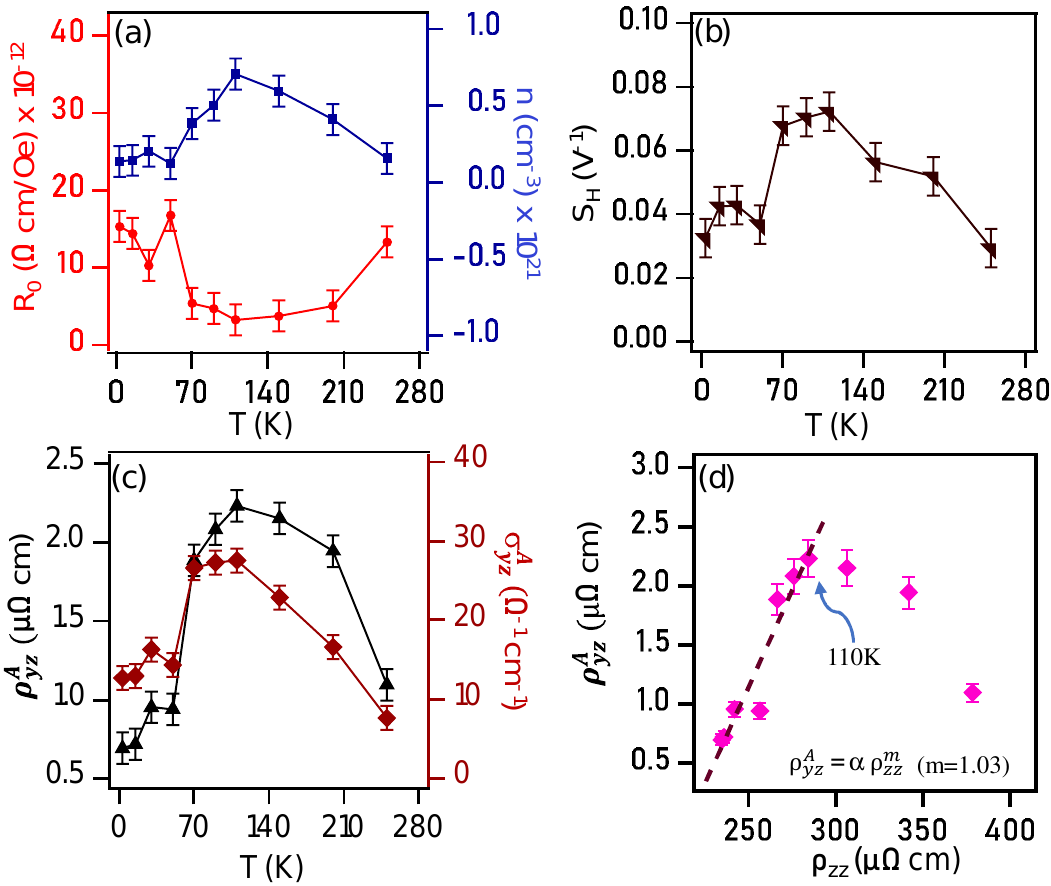}
    \caption{Temperature dependence of (a) Normal Hall coefficient $R_0$ (left axis) and charge carrier density $n$ (right axis). (b) Anomalous Hall scaling coefficient $S_H$ plotted as a function of temperature. (c) Anomalous Hall resistivity ($\rho^A_{\it{yz}}$) and Hall conductivity ($\sigma^A_{\it{yz}}$). (d) Plot of $\rho^A_{\it{yz}}$ $vs.$ $\rho_{\it{zz}}$. Dashed line in (d) is linear fitting with equation shown on the figure.}
    \label{4}
\end{figure}

\section{Results and Discussion}

Fig.~\ref{1}(a) shows the XRD pattern of Cr$_{2.76}$Te$_4$ single crystal with intensity peaks of $(l~0~0)$ Bragg plane, indicating that the crystal growth plane is along the $\it{a}$-axis. Inset in Fig.~\ref{1}(a) shows the photographic image of Cr$_{2.76}$Te$_4$ single crystal. Fig.~\ref{1}(b) shows XRD pattern of crushed Cr$_{2.76}$Te$_4$ single crystals measured at room temperature. All peaks in the XRD pattern can be attributed to the monoclinic crystal structure of $C12/m1$ space group (No.12) without any impurity phases, consistent with the crystal phase of Cr$_3$Te$_4$~\cite{Yamaguchi1972}. Rietveld refinement confirms the monoclinic structure with lattice parameters $\it{a}$=13.9655(2) \AA, $\it{b}$=3.9354(4)\AA, $\it{c}$=6.8651(7) \AA,
$\alpha$=$\beta$=90$^{\circ}$, and $\gamma$=118.326(7)$^{\circ}$. These values are in good agreement with previous reports on similar systems~\cite{doi:10.1143/JPSJ.54.1076,BABOT1973175}.  Fig.~\ref{1}(c) shows schematic crystal structure of  Cr$_{2.76}$Te$_4$  projected onto the $\it{ac}$-plane (top panel) and $\it{ab}$-plane (bottom panel). Cr1 atoms are located in the Cr-vacant layer with an occupancy of 0.189/$u.c$, whereas Cr2 atoms are located in the Cr-full layer with an occupancy of 0.5/$u.c$. The intercalated Cr atoms (Cr1) are sandwiched within the van der Waals gap created by the two CrTe$_2$ layers, as shown in Fig.~\ref{1}(c). The EDS measurements suggest an actual chemical composition of Cr$_{2.76}$Te$_4$.

To explore the magnetic properties of Cr$_{2.76}$Te$_4$, magnetization as a function of temperature [$M(T)$] was measured as shown in Fig.~\ref{2}(a) at a field of 1000 Oe applied parallel to the $\it{bc}$-plane ($H\parallel \it{bc}$) and $\it{a}$-axis ($H\parallel \it{a}$) for both zero-field-cooled (ZFC) and field-cooled (FC) modes. We observe that Cr$_{2.76}$Te$_4$ exhibits paramagnetic (PM) to ferromagnetic (FM) transition at a Curie temperature ($T_C$) of 310 K, which is close to the Curie temperature of Cr$_3$Te$_4$ ($T_C$=316 K).  Decrease in the sample temperature results into a decrease in the magnetization  for both $H\parallel \it{bc}$ and $H\parallel \it{a}$ at around 50 K, possibly due to spin-canting emerged from the coupling between in-plane ($\it{bc}$-plane) AFM order and out-of-plane ($\it{a}$-axis) FM orders~\cite{li2022diverse,PhysRevB.101.214413}. Also, the in-plane saturated magnetic moment of 1.78 $\mu_B$/Cr is almost 4 times higher than the out-of-plane saturated magnetic moment of 0.43 $\mu_B$/Cr at 2 K with an applied field of 1000 Oe, clearly demonstrating strong magnetocrystalline anisotropy in Cr$_{2.76}$Te$_4$. From the magnetization difference between ZFC and FC, $\Delta M=M_{FC} - M_{ZFC}$, shown in Fig.~\ref{2}(b), we notice significant magnetization fluctuations as the maximum of $\Delta M$ varies between in-plane and out-of-plane directions in going from 50 K to 150 K~\cite{Chen2021}.

The magnetic state of Cr$_{2.76}$Te$_4$ is further explored by measuring the magnetization isotherms [$M(H)$] for $H\parallel \it{a}$ and $H\parallel \it{bc}$ at various sample temperatures as shown in Figs.~\ref{2}(c) and \ref{2}(d), respectively. Consistent with $M(T)$ data, the magnetization saturation occurs at an applied field of 0.7 T and 1.4 T for $H\parallel \it{bc}$ and $H\parallel \it{a}$, respectively, suggesting $\it{bc}$-plane to be the easy-magnetization plane. Also, Cr$_{2.76}$Te$_4$ is a soft ferromagnet as it has a negligible coercivity [see Fig. S1(a) in the Supplemental Material~\cite{suppl}]. The observed saturation magnetisation ($M_s$) 2.586 $\mu_B$/Cr and 2.55 $\mu_B$/Cr for $H\parallel a$ and  $H\parallel bc$, respectively, are smaller than the stand alone  Cr atom (3 $\mu_B$), indicating correlated magnetic states in Cr$_{2.76}$Te$_4$. These observations are in good agreement with report on Cr$_{2.76}$Te$_4$~\cite{Shimada1996}.

\begin{figure*}[t]
    \centering
    \includegraphics[width=\linewidth]{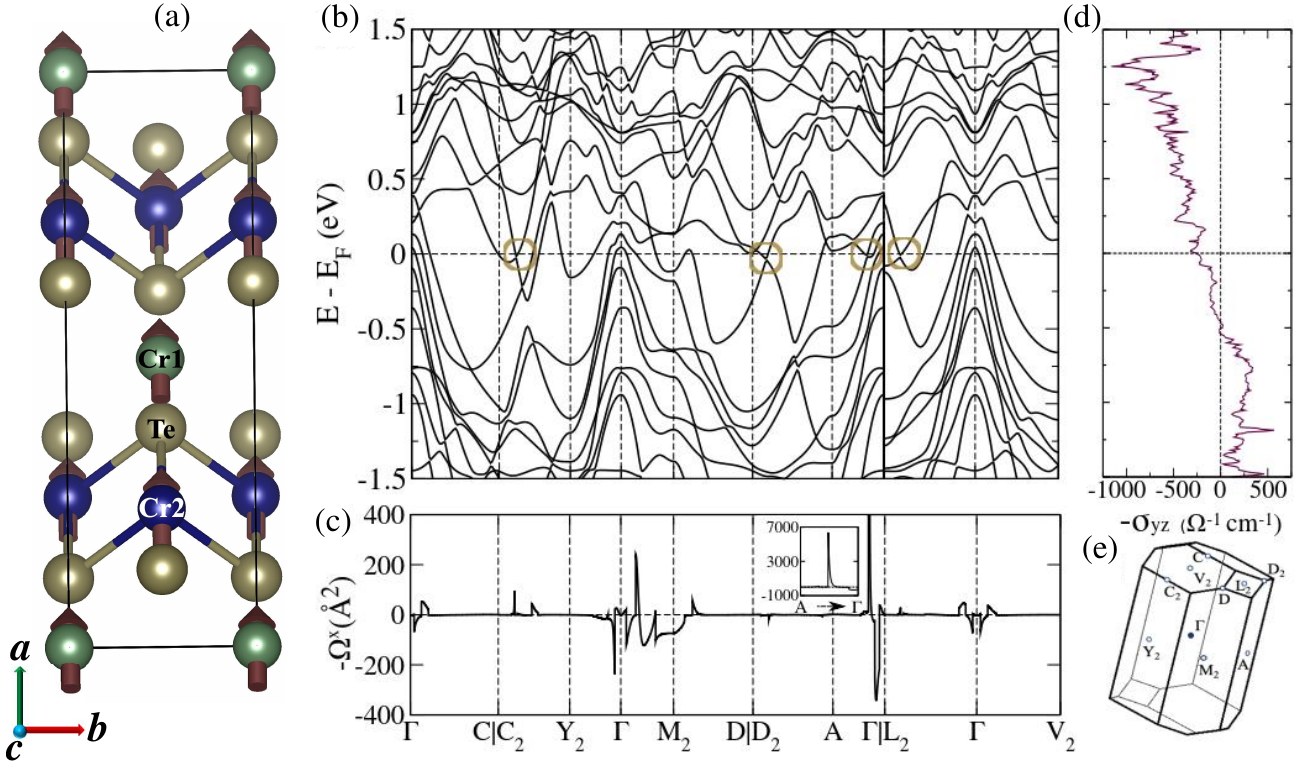}
    \caption{(a) Schematic monoclinic unit cell of Cr$_{2.76}$Te$_4$. (b) Electronic band structure of Cr$_{2.76}$Te$_4$ calculated with inclusion of spin-orbit coupling (SOC). In (b) the Weyl points near the Fermi level are encircled.   (c) $\it{x}$-component of the Berry curvature, $\Omega^{\it{x}}$,  calculated at the Fermi level. Inset in (c) shows zoomed-in view of $\Omega^{\it{x}}$ for the $A-\Gamma$ segment. (d) Anomalous Hall conductivity, $\sigma_{\it{yz}}$, plotted as a function of energy. (e) High symmetry points defined on the Brillouin zone of monoclinic primitive unit cell.}
    \label{6}
\end{figure*}

Next, coming to the main results of this contribution, Fig.~\ref{3}(a) exhibits temperature dependent longitudinal electrical resistivity ($\rho_{\it{zz}}$) of Cr$_{2.76}$Te$_4$. $\rho_{\it{zz}}(T)$ suggests metallic nature throughout the measured temperature range~\cite{PhysRevB.98.195122}. However, a $\it{kink}$ at around 50 K is noticed in the resistivity, related to the AFM ordering [see Fig.~\ref{2}(a)]. Bottom inset of Fig.~\ref{3}(a) depicts schematic diagram of linear-four-probe measuring geometry and the top inset elucidates the quadratic nature of low temperature resistivity up to 50 K as it can be explained well by the Fermi liquid (FL) theory,   $\rho(T)=\rho_0 + aT^2$ where $\rho_0$ is the temperature independent residual resistivity. Schematic diagram of Hall measuring geometry is shown in Fig.~\ref{3}(b). The Hall resistivity, $\rho_{\it{yz}}$,  is measured with current along the $\it{y}$-axis and magnetic field applied along the $\it{x}$-axis to get the Hall voltage along the $\it{z}$-axis. Thus, Fig.~\ref{3}(c) shows field dependent Hall resistivity $\rho_{\it{yz}}$ (black curve) measured at various sample temperatures. The total Hall resistivity ($\rho_{\it{yz}}$) may have contributions from the normal Hall effect ($\rho^N$) and the anomalous Hall effect ($\rho^A$). Thus, the total Hall resistivity can be expressed by the empirical formula, $\rho_{\it{yz}}(H)=\rho^N(H)+\rho^A(H)=\mu_0R_0H+\mu_0R_SM$, where $R_0$ and $R_S$ are the normal and anomalous Hall coefficients, respectively. These coefficients can be obtained by performing a linear fit using the relation $\frac{\rho_{yz}}{\mu_0H}=R_0+R_S\frac{M}{H}$ as shown in Fig. S1(c) of the Supplemental Material~\cite{suppl}. Having obtained the normal and anomalous Hall coefficients, we can now fit the total Hall resistivity (red curves) using the equation, $\rho_{\it{yz}}(H)=\mu_0R_0H+\mu_0R_sM$.  The fitting should be nearly perfect if there is no topological Hall contribution. However, from Fig.~\ref{3}(c) we can clearly notice that the fitting (red curve) is not perfect. Therefore, the topological Hall resistivity also contributes to the total Hall resistivity which can be expressed by $\rho_{\it{yz}}(H)=\rho^N(H)+\rho^A(H)+\rho^T(H)$. Thus, the topological Hall contribution (blue curve) is extracted using the relation, $\rho^T(H)= \rho_{\it{yz}}(H)-[\rho^N(H)+\rho^A(H)]$~\cite{Nakatsuji2015, Kuroda2017,Kanazawa2011}. See the Supplemental Material for more details~\cite{suppl}.

Fig.~\ref{4}(a) depicts the normal Hall coefficient ($R_0$) and the charge carrier density ($n$) derived using the formula, $(R_0=1/n|e|)$, plotted as a function of temperature. We clearly notice from Fig.~\ref{4}(a) that as the temperature decreases the carrier density increases up to 110 K. However, below 110 K,  the carrier density decreases with temperature and gets saturated below 50 K. Fig.~\ref{4}(b) (left axis) presents the anomalous Hall resistivity ($\rho^A_{\it{yz}}$) at zero field obtained by linearly intersecting the field axis [see Fig.~\ref{3}(c)]. Maximum anomalous Hall resistivity is noticed at around 110 K.  Anomalous Hall conductivity, $\sigma^{A}_{\it{yz}}$, derived using the formula $\sigma^{A}_{\it{yz}}=\frac{\rho_{\it{yz}}}{\rho^2_{\it{yz}}+\rho^2_{\it{zz}}}$ is shown in the right axis of Fig.~\ref{4}(c), again to find the maximum Hall conductivity of $\sigma^{A}_{\it{yz}}$=27 $\Omega^{-1} cm^{-1}$ at around 110 K. Fig.~\ref{4}(b) shows anomalous scaling coefficient $S_H=\frac{\rho^{A}_{yz}}{M\rho^{2}_{zz}}$, plotted as a function of temperature. The values of $S_H$ are inline with the itinerant ferromagnetic systems of $S_H$=0.01-0.2 $V^{-1}$~\cite{Kaul1979}. In general, anomalous Hall effect can occur in solids either intrinsically originated from the nonzero-Berry curvature in the momentum space~\cite{RevModPhys.82.1539,PhysRevLett.96.037204} or extrinsically due to side-jump/scew-scattering mechanisms~\cite{Smit1958, RevModPhys.82.1539,PhysRevB.2.4559}. Therefore, to elucidate the mechanism of AHE in Cr$_{2.76}$Te$_4$, we plotted $\rho^{A}_{\it{yz}}$ $vs.$ $\rho_{\it{zz}}$ as shown in Fig.~\ref{4}(d). From Fig.~\ref{4}(d) it is evident that $\rho^A_{\it{yz}}$ linearly changes with $\rho_{\it{zz}}$ [$\rho_{\it{yz}}=\alpha \rho_{\it{zz}}^m$ for $m=1.03\pm0.02$] up to 110 K and then deviates on further increasing the sample temperature~\cite{Shen2022}.   In the case of itinerant ferromagnetic system, the Hall resistivity can be expressed by the relation $\rho_{\it{yz}}=\alpha \rho_{\it{zz}}+\beta \rho_{\it{zz}}^2$ where $\alpha$ and $\beta$ are the screw-scattering and side-jump terms, respectively~\cite{PhysRevLett.103.087206,PhysRevB.98.195122,PhysRevB.96.134428}. That means, in the case of skew-scattering $\rho^{A}_{\it{yz}}$ linearly depends on $\rho_{\it{zz}}$, while $\rho^{A}_{\it{yz}}$  quadratically depends on $\rho_{\it{zz}}$ in the case of side-jump.  Since the Hall resistivity ($\rho_{\it{yz}}$) linearly depends on $\rho_{\it{zz}}$, the skew-scattering could be the most suitable mechanism of AHE observed in this system~\cite{Liu2018, Jiang2020}. Note that the intrinsic Berry curvature contribution to the AHR also quadratically dependents on $\rho_{\it{zz}}$~\cite{Shen2022}. To make a point, it looks that 110 K is the critical temperature of the properties discussed in Figs.~\ref{4}(a)-(d). However, as per the $dM/dT$ data shown in the inset of Fig.~\ref{2}(b), we think that the critical temperature could be $\approx$ 125 K instead of 110 K as we can see significant change in magnetization across $\approx$ 125 K.

\begin{figure}[t]
    \centering
    \includegraphics[width=\linewidth]{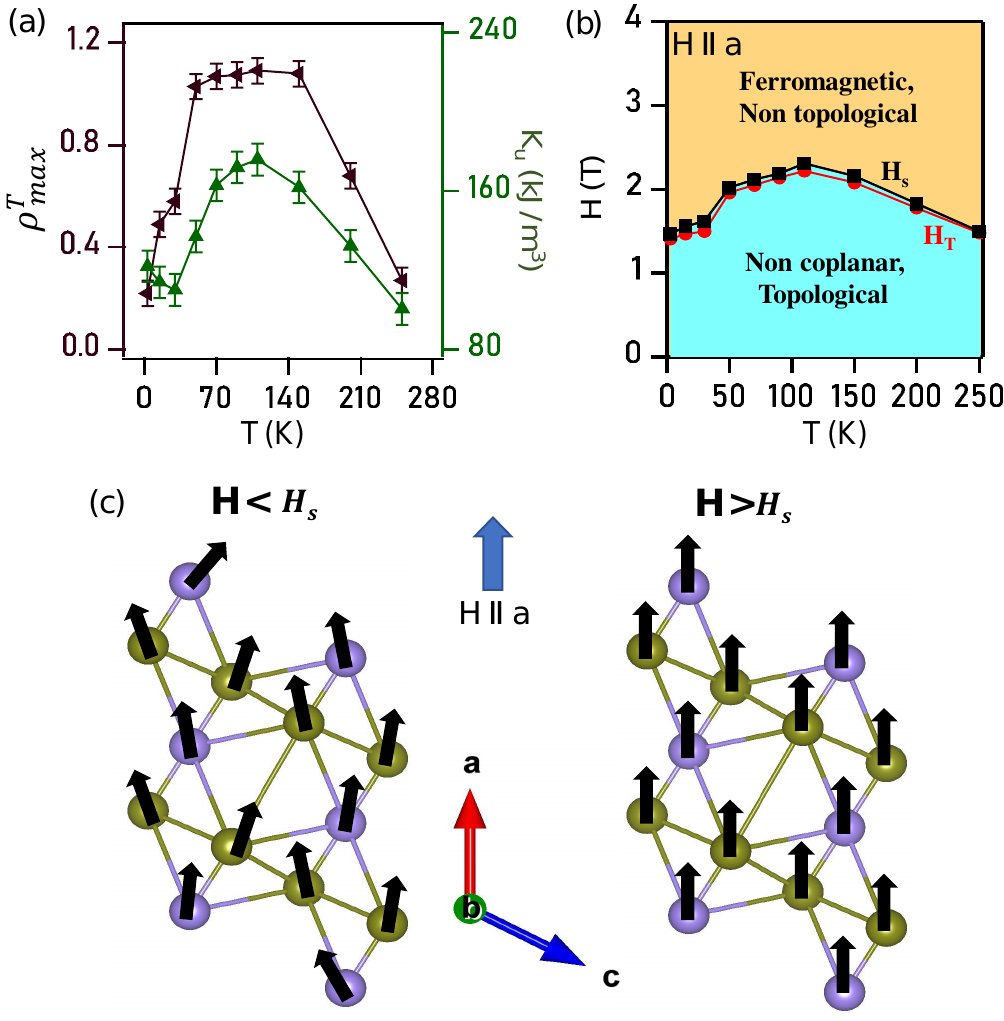}
    \caption{(a) Maximum value of $\rho^T_{max}$ and magnetocrystalline anisotropy ($K_u$) plotted as a function of temperature. (b) Saturation magnetic field ($H_s$) and magnetic field ($H_T$) above which the topological Hall effect disappears for $H\parallel \it{a}$.   (c) Schematic representation of $Cr$  spin configuration for $H<H_s$ (left) and  $H>H_s$. See the text for more details. }
    \label{5}
\end{figure}

Our experimental findings on the AHE have been examined using the density functional theory calculations as presented in Fig.~\ref{6}. For the calculations, we considered primitive unit cell of Cr$_{3}$Te$_4$, consisting of three Cr (one Cr1 and two Cr2 type) and four Te atoms. The magnetic spins of Cr atoms were considered along the $\it{x}$-direction. Our calculations suggest a ferromagnetic ground state with average magnetic moments of 3.23$\mu_B$ and 3.15$\mu_B$ per Cr1 and Cr2 atoms, respectively, slightly higher than the experimental average value of 2.568 $\mu_B/Cr$. In Fig.~\ref{6}(b), we present bulk electronic band structure of Cr$_{3}$Te$_4$. The system is found to be metallic with several bands crossing the Fermi level ($E_F$). Several band crossing points are found near $E_F$ along $C_2-Y_2$, $D_2-A$, $A-\Gamma$ and $L_2-\Gamma$ $k$-paths, but slightly away from the high symmetry points.  Next, we explore the Berry curvature ($\Omega$) calculated using the formula $\Omega(k)= \nabla(k)\times A(k)$ (where $A(k)$ is the Berry connection). Variation of the $\it{x}$-component of Berry curvature ($\Omega^x$) at $E_F$ along the high symmetry $k$-path is shown in Fig.~\ref{6}(c). We can notice from Fig.~\ref{6}(c) that $\Omega^x$ is strongly enhanced along the $A-\Gamma$ ($k_z$)direction. Further, Fig.~\ref{6}(d) shows the anomalous Hall conductivity (AHC), $\sigma_{\it{yz}}$,  of Cr$_{3}$Te$_4$ calculated as a function of energy using the Eqn.~\ref{eq1}.

\begin{equation}\label{eq1}
\sigma_{\it{yz}}=\frac{e^2}{\hbar} \int f(\epsilon_n(\textbf{k})) \Omega_{n}^{\it{x}}(\textbf{k})\frac{d\textbf{k}}{(2\pi)^3}
\end{equation}
Where, $f(\epsilon_n(\textbf{k}))$ is  the Fermi-Dirac distribution function.

 Our calculations suggest an intrinsic AHC of $\sigma_{\it{yz}}$$\approx$260 $\Omega^{-1}~cm^{-1}$ near $E_F$, much larger than the experimental value of $\sigma^A_{\it{yz}}$=27 $\Omega^{-1}~cm^{-1}$. Such a small AHC suggests dominant impurity scattering contribution to the total anomalous Hall effect of Cr$_{3}$Te$_4$~\cite{Liu2021}, i.e., the skew-scattering contribution in the dirty limit~\cite{PhysRevLett.97.126602}. However, despite the system is in the dirty limit,  the total AHC should be at least comparable to the value of intrinsic AHC ($\approx$260 $\Omega^{-1}~cm^{-1}$) as it has contributions from both intrinsic Berry-curvature and extrinsic skew-scattering. In contrast, experimentally, we find a much smaller AHC compared to theoretical calculations. Note here that the calculations performed on the stoichiometric Cr$_3$Te$_4$ predict a large intrinsic AHC near $E_F$ due to the presence of several band crossing points (Weyl points). But the intrinsic AHC rapidly decreases as we move away from $E_F$. This is particularly true when $E_F$ is shifted to lower binding energies [see Fig.~\ref{6}(d)].  Therefore, a genuine reason behind the discrepancy of AHC between experiment and theory could be that the experiments are performed on a slightly off-stoichiometric composition of Cr$_{2.76}$Te$_4$ with 8\% Cr deficiency per formula unit, shifting the Fermi level towards the lower binding energy. Maximum topological Hall resistivity ($\rho^T_{max}$) is plotted as a function of temperature in Fig.~\ref{5}(a) by the black-colored data points. Also, green-colored data points in Fig.~\ref{5}(a) represent the magnetocrystalline anisotropy constant $K_u$ calculated using the Eqn.~\ref{eq2}.

\begin{equation}\label{eq2}
K_u=\mu_0\int_{0}^{M_s}[H_{bc}(M)-H_{a}(M)]~dM \\
\end{equation}
Here, M$_s$ represents saturation magnetization.  H$_{bc}$ and H$_{a}$ represent $H\parallel \it{bc}$ and $H\parallel \it{a}$, respectively.

From Fig.~\ref{5}(a), we can notice that the topological Hall resistivity is highest within the temperature range of 50 and 150 K. Also, most importantly,  the temperature dependance of $K_u$ resembles $\rho^T_{max}$. In Fig.~\ref{5}(b), the black-colored data points illustrate temperature dependent saturation field ($H_s$) beyond which the system becomes ferromagnetic [extracted from Fig.~\ref{2}(c)] and the red-colored data points illustrate temperature dependent field ($H_{T}$) beyond which $\rho^T_{\it{yz}}$  becomes zero [extracted from Fig.~\ref{3}(c)]. Interestingly, both fields $H_{T}$ and $H_s$ perfectly overlap at all measured temperatures. This can be understood using the schematics shown in Fig.~\ref{5}(c). Means, at a given temperature up to the saturation field ($H<H_s$),  the system possess non-coplanar spin structure and hence shows the topological Hall effect~\cite{zheng2021giant}. However, for the applied field beyond magnetic saturation ($H>H_s$), the system becomes ferromagnetic and thus THE disappears.

Further, the maximum topological Hall resistivity $\rho^T_{max}$ $\approx1.1$ $\mu\Omega$-cm over a broad temperature range of $(50K < T < 150K)$ observed in this vdW ferromagnetic Cr$_{2.76}$Te$_4$ is in the same order of magnitude found in the chiral semimetals such as Mn$_2$PtSn~\cite{liu2018giant}, Gd$_3$PdSi$_3$~\cite{kurumaji2019skyrmion},  LaMn$_2$Ge$_2$~\cite{PhysRevMaterials.5.034405}, and in other Cr$_x$Te$_y$ based systems~\cite{PhysRevB.100.024434,huang2021possible}. Interestingly, so far THE is observed in Cr$_x$Te$_y$ systems in their hexagonal (trigonal) crystal structure~\cite{he2020large, PhysRevB.100.024434,huang2021possible} but not in the monoclinic structure that we found in the present study.    As discussed above, the easy-axis of magnetization in Cr$_{2.76}$Te$_4$ is found to be in the $\it{bc}$-plane and thus for small applied fields the Cr spins are canted out of the $\it{bc}$-plane towards the $\it{a}$-axis for $H\parallel \it{a}$. In this way, the non-coplanar spin-structure has been generated for sufficiently smaller fields~\cite{he2020large, li2020large,PhysRevMaterials.5.034405,wang2020giant, PhysRevB.100.024434,huang2021possible}.

Several mechanisms are proposed to understand the topological Hall effect. Such as the antisymmetric exchange or Dzyaloshinskii–Moriya (DM) interaction in the noncentrosymmetric systems~\cite{yu2010real,jiang2017skyrmions,PhysRevLett.113.087203} and the uniaxial magnetocrystalline anisotropy in the centrosymmetric systems~\cite{PhysRevB.99.094430,ding2019observation,gilbert2015realization,doi:10.7566/JPSJ.87.074704}. In the case of monoclinic Cr$_{3}$Te$_4$ ($C21/m1$) which is a centrosymmetric crystal, the chiral-spin structure could be stabilized by the strong MCA. This analogy is completely supported by our experimentally estimated MCA values at various temperatures as shown in Fig.~\ref{5}(a). Most importantly, highest $\rho^T_{max}$ value has been obtained at the highest magnetocrystalline anisotropy of $K_u$=165 $kJ/m^3$. This is because, in presence of the chiral-spin structure,  the itinerant electrons acquire real-space Berry curvature associated with finite scalar-spin chirality  $\chi_{ijk}=S_{i}.(S_{j}\times S_{k})$ which serves as fictitious magnetic field to generate the topological Hall signal~\cite{PhysRevLett.98.057203,taguchi2001spin,wang2022topological}.

\section{Conclusions}
To summarize, we have grown high quality single crystal of layered ferromagnetic Cr$_{2.76}$Te$_4$ in the monoclinic phase to study the electrical transport, Hall effect, and magnetic properties. Our studies suggest Cr$_{2.76}$Te$_4$ to be a soft ferromagnet with a negligible coercivity. The easy-axis of magnetization is found to be parallel to the $bc$-plane, leading to strong magnetocrystalline anisotropy. Below 50 K, an antiferromagnetic-like transition is noticed. Interestingly, in going from 50 K to 150 K the strength of magnetic moments switches between out-of-plane to in-plane, suggests fluctuating Cr spins. From the electrical resistivity measurements the system is found to be metallic throughout the measured temperature range. Also, a $kink$ at around 50 K due to AFM ordering is noticed. Magnetotransport  measurements demonstrate large anomalous Hall effect (AHE) and topological Hall effect (THE) in this systems. First-principles calculations point to an intrinsic AHE due to non-zero Berry curvature near the Fermi level, while experimentally it is found to be an extrinsic AHE due to skew-scattering. Topological Hall effect has been observed due to the non-coplanar spin-structure in the presence of strong magnetocrystalline anisotropy.

\section{Acknowledgements}
A.B. acknowledges support from Prime Minister's Research Fellowship (PMRF). A.N. thanks startup grant of the Indian Institute of Science (SG/MHRD-19-0001) and DST-SERB (SRG/2020/000153). S.T. thanks the Science and Engineering Research Board (SERB), Department of Science and Technology (DST), India for the financial support (Grant No.SRG/2020/000393). This research has made use of the Technical Research Centre (TRC) Instrument Facilities of S. N. Bose National Centre for Basic Sciences, established under the TRC project of Department of Science and Technology, Govt. of India.
\bibliography{Cr3Te4}
\end{document}